%% file: paper.tex
\documentclass{article}

\usepackage[final,nonatbib]{neurips_2023_ml4ps} %[]final

\usepackage[utf8]{inputenc} % allow utf-8 input
\usepackage[T1]{fontenc}    % use 8-bit T1 fonts

\usepackage[sorting=none,style=numeric-comp]{biblatex}
\usepackage{hyperref}       % hyperlinks
\usepackage{url}            % simple URL typesetting
\usepackage{booktabs}       % professional-quality tables
\usepackage{amsfonts}       % blackboard math symbols
\usepackage{nicefrac}       % compact symbols for 1/2, etc.
\usepackage{microtype}      % microtypography
\usepackage{xcolor}         % colors

\usepackage{amsmath,amssymb,amsthm}
\usepackage{mathtools}
\usepackage{graphicx}
\usepackage{subcaption}
\usepackage{listings}
\usepackage{placeins}
\usepackage{xspace}
\usepackage{todonotes}

\newcommand{\pt}{\ensuremath{{p}_{\mathrm{T}}}\xspace}
\newcommand{\rel}{\ensuremath{\mathrm{rel}}}

\newcommand{\etarel}{\ensuremath{\eta^{\rel}}\xspace}
\newcommand{\phirel}{\ensuremath{\phi^{\rel}}\xspace}
\newcommand{\mrel}{\ensuremath{m^{\rel}}\xspace}

\newcommand{\ptjet}{\ensuremath{{p}_{\mathrm{T,jet}}}\xspace}
\newcommand{\ptrel}{\ensuremath{{p}_{\mathrm{T}}^{\rel}}\xspace}

\newcommand{\JN}{\texttt{JetNet}\xspace}

\newcommand{\TeV}{\ensuremath{\,\text{Te\hspace{-.08em}V}}\xspace}

\usepackage{xcolor}
\definecolor{msg}{rgb}{0.59,0.45,0.65}
\definecolor{ngbr}{rgb}{0.52,0.42,0.0}
\definecolor{x4}{rgb}{0.51,0.70,0.40}

\title{DeepTreeGANv2: Iterative Pooling of Point Clouds}

\usepackage{orcidlink}

\author{
  Moritz A.W. Scham\,\orcidlink{0000-0001-9494-2151}\thanks{Also at J\"ulich Supercomputing Centre, Institute for Advanced Simulation, Germany and RWTH Aachen University, III. Physikalisches Institut A, Germany}\\
  Deutsches Elektronen-Synchrotron DESY,\\
  Germany \\
  \texttt{moritz.scham@desy.de} \\
  \AND
  Dirk Krücker\,\orcidlink{0000-0003-1610-8844}\\
  Deutsches Elektronen-Synchrotron DESY,\\
  Germany \\
  \texttt{dirk.kruecker@desy.de} \\
  \AND
  Kerstin Borras\,\orcidlink{0000-0003-1111-249X}\thanks{Also at RWTH Aachen University, III. Physikalisches Institut A, Germany}\\
  Deutsches Elektronen-Synchrotron DESY,\\
  Germany \\
  \texttt{kerstin.borras@desy.de} \\
}

\begin{document}

\maketitle

\begin{abstract}
In High Energy Physics, detailed and time-consuming simulations are used for particle interactions with detectors. To bypass these simulations with a generative model, the generation of large point clouds in a short time is required, while the complex dependencies between the particles must be correctly modelled. Particle showers are inherently tree-based processes, as each particle is produced by the decay or detector interaction of a particle of the previous generation.
In this work, we present a significant extension to DeepTreeGAN~\cite{DeepTreeCHEP}, featuring a critic, that is able to aggregate such point clouds iteratively in a tree-based manner. We show that this model can reproduce complex distributions, and we evaluate its performance on the public JetNet 150 dataset.
\end{abstract}

\input{parts/intro}
\input{parts/architecture}
\input{parts/results}

\section{Conclusion}
In this work, the DeepTreeGAN model, first introduced in~\cite{DeepTreeCHEP}, was significantly extended to model up to 150 jet constituents of the \JN dataset. To this end, a PC-based critic has been developed, featuring a pooling operation (Section~\ref{sec:pooling}), that allows the iterative downscaling of PCs. Through its significant advantages, this novel pooling implementation may prove useful not only for generative tasks, but also any PC regression or classification task.
This extension represents a significant milestone in scaling this model to handle even larger point clouds, such as particle showers in high-granularity calorimeters.
% This paper presents a new model (DeepTreeGAN) for generating large point clouds using a novel upscaling method for point clouds. It successfully replicates the complex distributions of the \JN datasets with high fidelity (Section~\ref{sec:resul}) and demonstrating its suitability for generating high-dimensional point clouds. The model's tree-based structure promises good scaling behavior for even larger point clouds, such as particle showers in high-granularity calorimeters. However, the preliminary version presented here uses the MPGAN critic that scales with $\mathcal{O}(\mathrm{points}^2)$ and thus, requires a new critic with better scaling behavior. In addition, the generator will need to be able to generate a variable number of points for each event, especially for the \JN datasets with up to 150 particles and the calorimeter use case. This will be the subject of future publications.

% \section{Limitations}
% slow convergence 300 hour V100
% number of points input of the model
% high memory usage
% compute heavy training 

\section*{Acknowledgement}
Moritz Scham is funded by Helmholtz Association's Initiative and Networking Fund through Helmholtz AI (grant number: ZT-I-PF-5-3).
This research was supported in part through the Maxwell computational resources operated at Deutsches Elektronen-Synchrotron DESY (Hamburg, Germany). The authors acknowledge support from Deutsches Elektronen-Synchrotron DESY (Hamburg, Germany), a member of the Helmholtz Association HGF.

% \clearpage
%  \typeout{}
% \bibliography{references}

\printbibliography
\end{document}

%% file: parts/intro.tex
\section{Introduction}\label{intro}

In particle physics, detailed, near-perfect simulations of the underlying physical processes, the measurements, and the details of the experimental apparatus are state-of-the-art. However, accurate simulations of large and complex detectors, especially calorimeters, using current Monte Carlo-based tools such as those implemented in the Geant4~\cite{Agostinelli2003} toolkit
are computationally intensive. Currently, more than half of the worldwide Large Hadron Collider
(LHC) grid resources are used for the generation and processing of simulated data~\cite{Albrecht2019}. 
For the future High-Luminosity upgrade of the LHC (HL-LHC~\cite{Apollinari2015}), the computational requirements will exceed the computational resources without significant speedups, e.g., for the CMS experiment by 2028~\cite{Software2022} projecting the current technologies.
This will become even more demanding for future high-granularity calorimeters, e.g., the CMS HGCAL~\cite{CMS2017} with its complex geometry and extremely large number of channels.
To address these challenges, fast simulation approaches based on Deep Learning, including Generative Adversarial Networks (GAN)~\cite{Goodfellow2020} and Variational Autoencoders (VAE)~\cite{Kingma2013}, have been explored early~\cite{Oliveira2017,Paganini2018,Paganini2018a,Erdmann2018,Erdmann2019,Carminati2018} and deployed recently~\cite{Aad2022}.

The majority of these approaches consider the data as voxels living on three-dimensional grids. Especially for high-granularity calorimeters like the HGCAL, the calorimeter cells are highly irregular, and the data is often sparse. In such a situation, it is beneficial to consider the data as Point Clouds (PC), e.g., tuples of cell energies and three-dimensional coordinates, i.e., four-dimensional point clouds~\cite{kaech2022jetflow,kaech2022point,schnake2022}. For modelling calorimeter showers, which are cascades of particles, it seems natural to model the creation of such point clouds as a tree-like process. Such an approach can be implemented by Graph Neural Networks (GNN)~\cite{Scarselli2009} and Graph Convolutions~\cite{kipf2016}. For modelling PC data representing three-dimensional objects, a similar approach~\cite{Shu2019} called tree-GAN~\cite{Shu2019} exists. For our use case, especially for modelling the large dynamic range of the energy dimension, this approach is not sufficient.
Here we report on the development of a largely extended Graph GAN approach, which we named {\it DeepTreeGAN} that can be applied to multiple areas of PC generation. To demonstrate the abilities of our model, we present a first application to the \JN dataset introduced in the next section. 

\subsection{JetNet}\label{sec:jetnet}

In this study, the \JN~\cite{JetNet30v2,JetNet150v2} datasets are used. 
Each dataset contains PYTHIA~\cite{Sjostrand2008} jets with an energy of about 1\TeV, with each jet containing up to 30 or 150 constituents (here: 150). The datasets differ in the jet-initiating parton. Here, the dataset for top quark %, light quarks and gluon-initiated 
jets is studied. Each dataset contains about 170k individual jets split as 110k/10k/50k for training/test/validation, where the validation dataset is used for our results. The jet constituents, the particles, are clustered with a cone radius of $R = 0.8$. These particles are considered to be massless and can therefore be described by their 3-momenta or equivalently by transverse momentum $p_T$, pseudorapidity $\eta$, and azimuthal angle $\phi$. In the \JN datasets, these variables are given relative to the jet momentum: 
$\etarel_i \coloneqq \eta_i - \eta^\mathrm{jet}$,
$\phirel_i \coloneqq \phi_i - (\phi^\mathrm{jet} \mod 2\pi)$, and %\fxfatal{BLAH}
$p_{T,i}^\mathrm{rel} \coloneqq p_{T,i}/\ptjet$, where $i$ runs over the particles in a jet. 
Calculating the invariant mass from these relative quantities, for example, for the jet mass, implies
%$\mreljet = m_{jet}/\ptjet$. 
$\mrel = m_{jet}/\ptjet$. 
This dataset has gained popularity in the particle physics community as a benchmark for PC-based generative models~\cite{Buhmann2020,Buhmann2021a,schnake2022,kaech2022jetflow,kaech2022point,kaech2023mdma,Mikuni2022,Mikuni2023,Kansal2022,Kansal2023,Leigh2023,Buhmann2023,buhmann2023epicly}.

%% file: parts/architecture.tex
\section{Architecture}\label{sec:arch}
PC-based GANs ~\cite{kaech2022point,kaech2023mdma,Kansal2022,Buhmann2023} frequently choose a refinement approach:
Starting with a PC of the desired size with features sampled from noise, the PC is updated multiple times in the generator and the critic. In the final step of the critic, the points are aggregated and mapped to a single value, usually with a feedforward neural network (FFN). In this way, the dimensionality is greatly reduced in a single step.
Commonly, the PC is updated by first transforming the points individually. Then they are aggregated into a global vector, which is used in-turn to transform the points again.
In an image-based GAN approach like DCGAN~\cite{dcgan}, the generator would apply convolutions in sequence to iteratively upscale a random vector to an image and the critic would apply convolutions in sequence to iteratively downscale an image to a scalar.
These convolutions cannot be directly translated to PCs, which is why the mentioned refinement approach is prevalent for PC-based GANs.
Our approach instead is to translate this principle and to construct a critic that reduces the dimensionality iteratively.
In this way, we complement the DeepTree generator \cite{DeepTreeCHEP}, that iteratively upscales PCs.

\subsection{Generator}
The generator part of this model largely matches the one published in~\cite{DeepTreeCHEP}, with a few key differences: 
Instead of splitting the leaf vector into one part per branch and mapping the latter separately to the desired number of branches, the full leaf vector is now mapped to the number of branches times number of features.
The generator uses 2, 3, 5, and 5 branches (with a product of 150) and goes from 64 to 33 to 20 to 10 to 3 features.
For each event, the number of constituents is sampled from the dataset and the output of the generator is cut to this size.

\subsection{Critic}\label{sec:critic}
\begin{figure}
    \centering
    \includegraphics[width=0.75\textwidth]{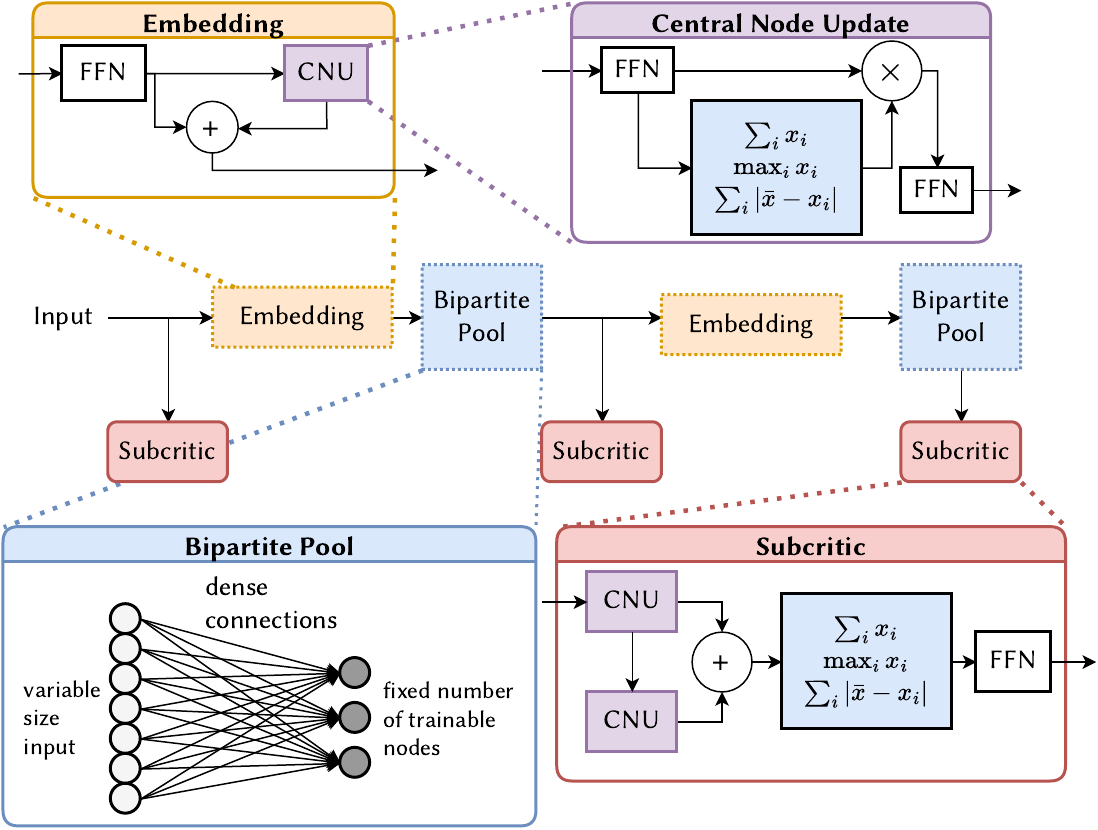}
    \caption{The critic, as described in \ref{sec:critic}}
    \label{fig:critic}
\end{figure}

The critic is shown in Figure~\ref{fig:critic}, its \emph{components} are described in the following section. It features three \emph{subcritics}, that are applied to the different stages of aggregation: The first is applied directly on the input PC (up to 150 points), the other two are applied after each \emph{bipartite pool} (pooling to 30 points and 6 points).
The expectation is that the first \emph{subcritic} provides feedback on the more low-level features, while the two later \emph{subcritics} operate on aggregated points and thus provide feedback on more high-level features.
Before the \emph{bipartite pool}, the \emph{embedding} layer is applied, which maps the number of features to 10 and transforms the input.\\
The critic is conditioned on $\pt, \eta$ and the mass of the jet.

\paragraph{Bipartite Pool}\label{sec:pooling}
For the iterative reduction of the number of points a pooling operation should fulfill the following conditions: First, it must be differentiable to enable backpropagation. Then it should be invariant under permutation because the input is permutation invariant. As the number of input points varies, it needs to accept an arbitrary number of points, down to a single one. The run-time pooling operation should scale well to large PCs, ideally linear with the number of points. Finally, it should map to a fixed number of points to construct a regular shaped output tensor, that allows faster and more convenient implementations.
To achieve these requirements, we propose constructing a bipartite graph, densely connecting the input PC to a fixed number of trainable nodes and applying a Message Passing Layer (MPL) to this graph. This will return a PC with a fixed number of points.
The dense connection yields a number of edges that equals the number of input points times the number of trainable nodes.
Because each edge yields one message, the run-time of the MPL should be proportional to the size of the input PC.
As an MPL, Gatv2Conv~\cite{gatv2conv} is used with 16 attention heads, as implemented in PyTorch Geometric~\cite{pyg}.\\
A similar approach has been presented in \cite{Lee2019}, but implemented as an attention mechanism instead of an MPL. An advantage of our choice is that the bipartite pool can handle variable sized PCs without masking.

\paragraph{FFNs}
The \emph{FFNs} used in this critic consist of 3 layers with 100 hidden nodes without bias. 
The first two hidden layers are followed by a 50\% dropout layer and LeakyReLU activation with a negative slope of 0.1. Spectral normalization~\cite{Miyato2018} is applied to the \emph{FFNs}, except for the \emph{FFN} in the \emph{embedding} layer, where batch normalization~\cite{batchnorm} is applied before the activation.

\paragraph{Multi-Aggregation}
At different stages, the points need to be aggregated in a single vector. This is commonly done by summing over the points, e.g., \cite{Buhmann2023}. To provide additional information about the distribution,  we compute this vector by concatenating the sum, the maximum, the number of points and the width. The width of the distribution is estimated by computing the mean absolute deviation: $\frac{1}{n}\sum_i |x_i - \bar{x}|$.

\paragraph{Central Node Update}
For the following description of the \emph{embedding} layer and the \emph{subcritics}, we define a \emph{Central Node Update} Layer (\emph{CNU}): First, the input is transformed with an \emph{FFN}.
Then the transformed PC is aggregated with the \emph{multi-aggregation}.
Lastly, each of the transformed nodes is concatenated with the aggregated vector and passed through a second \emph{FFN} that maps the nodes back to their original dimension.

\paragraph{Embedding}
The \emph{embedding} layer first maps the points independently to 10 features with an \emph{FFN} and passes the PC though a \emph{CNU} layer with a residual connection.

\paragraph{Subcritics}
Each subcritic is constructed using two \emph{CNUs} with a residual connection. The points are then aggregated in the \emph{multi-aggregation} scheme. This aggregated vector is then concatenated with the condition ($\pt,\eta$ and mass of the jet) and passed to an \emph{FFN}, mapping the vector to a single output.
This yields three output values for each event which are summed such that the application of the loss and the backpropagation takes place for all the outputs in parallel.

\subsection{Training}
Generator and critic are trained using the Hinge~\cite{Hinge} loss, with a ratio of 1 to 2 gradients steps.
The optimizer for the generator (critic) is Adam~\cite{Kingma2015} with $\beta_1 = 0.9,\beta_2 = 0.999$, a weight decay of $10^{-4}$ and a learning rate of $10^{-5}$ ($3*10^{-5}$).
Additional to the Hinge loss, the generator is also trained to minimize the feature matching loss \cite{salimans2016improved} with a factor of 0.1. As an ``intermediate layer'' the concatenation of the \emph{multi-aggregation} of the input PC, together with the output of the \emph{embedding} and \emph{bipartite pool\textbf{}} layers are used.

\subsection{Preprocessing \& Postprocessing}
For the training, the $\etarel$ and $\phirel$ distributions are scaled separately to a normal distribution. The $\ptrel$ distribution is scaled using a Box-Cox transformation with standardizing as implemented in~\cite{Sklearn}.
To produce samples, the inverse scaling is applied to the output of the generator.
The original dataset is scaled so that $\sum_i \pt^i = 1$. This feature is difficult to model for the generator, while it is easy to recognize for the critic. To avoid limiting the performance of the model, the output of the generator is scaled in the same way.\\

%% file: parts/results.tex
\section{Results}\label{sec:resul}
\begin{figure}[htb]
    \centering
    \begin{subfigure}[b]{0.245\textwidth}
        \centering
        \includegraphics[keepaspectratio,width=.99\textwidth,trim={0.3cm 0.5cm 0.2cm 0.25cm},clip]{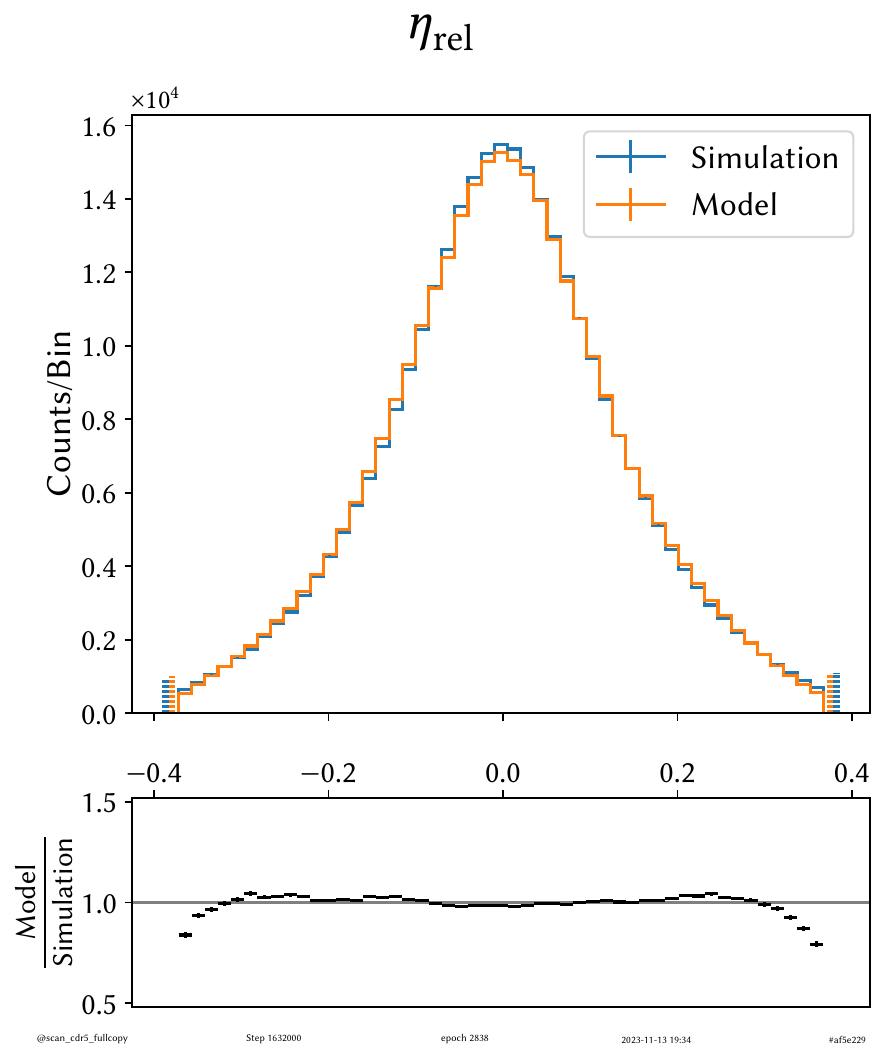}
    \end{subfigure}
    \begin{subfigure}[b]{0.245\textwidth}
        \centering
        \includegraphics[keepaspectratio,width=.99\textwidth,trim={0.3cm 0.5cm 0.2cm 0.25cm},clip]{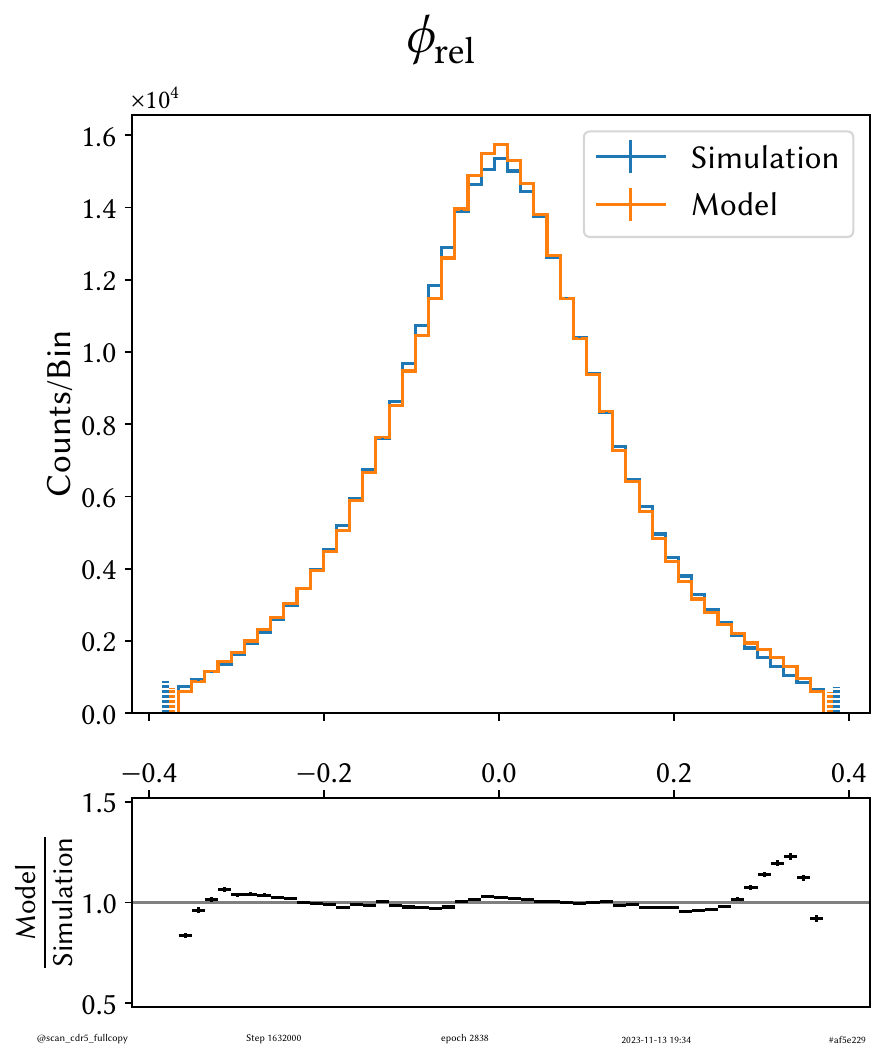}
    \end{subfigure}
    \begin{subfigure}[b]{0.245\textwidth}
        \centering
        \includegraphics[keepaspectratio,width=.99\textwidth,trim={0.3cm 0.5cm 0.2cm 0.25cm},clip]{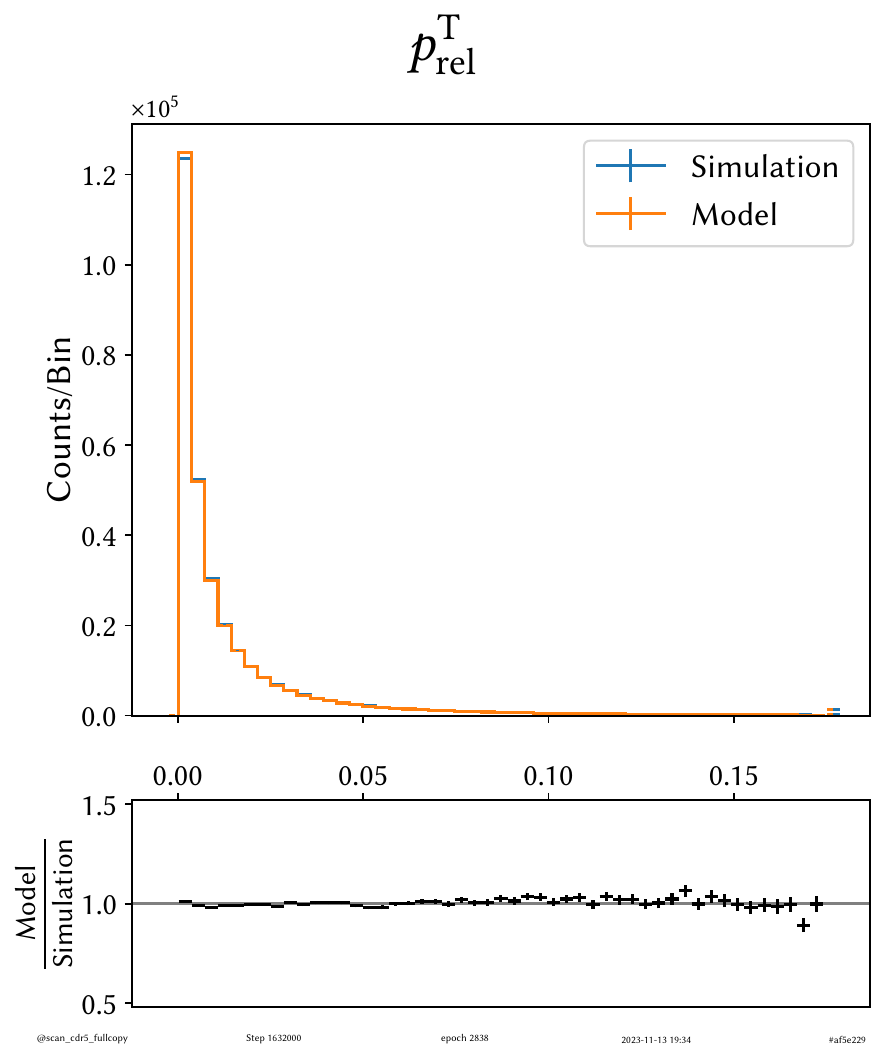}
    \end{subfigure}
    \begin{subfigure}[b]{0.245\textwidth}
        \centering
        \includegraphics[keepaspectratio,width=.99\textwidth,trim={0.3cm 0.5cm 0.2cm 0.25cm},clip]{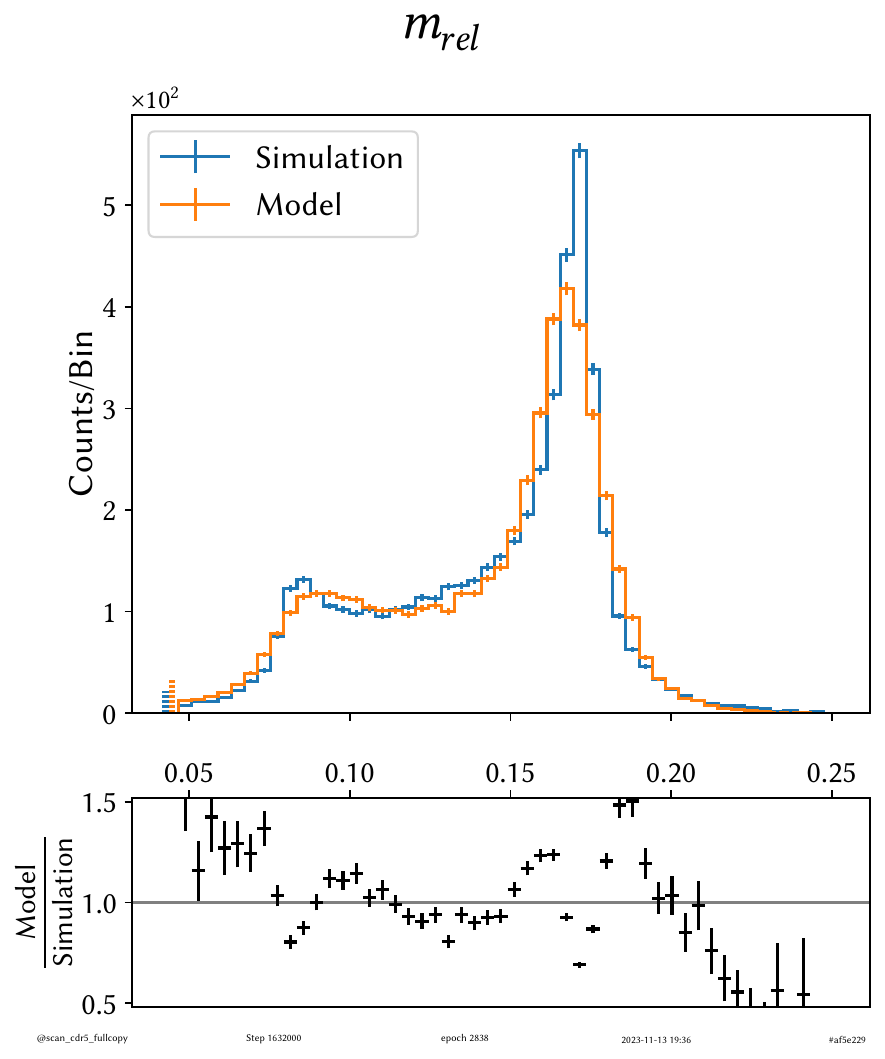}
    \end{subfigure}
    \caption{The distributions of $\etarel, \phirel, \ptrel$ for constituents of the jets and the mass of the jets for the top-quarks \JN 150 dataset (Simulation), as described in Section~\ref{sec:jetnet} and DeepTreeGAN (Model).}\label{fig:t-mar}
\end{figure}
\input{parts/metricstable}
The variables of the constituents and the mass of the top quark dataset are shown in Figure~\ref{fig:t-mar}. 
The generated distributions of the constituents closely match the distribution of the data. While the double peak structure of the mass distribution is clearly visible, the model is not able to produce the peak quite as sharp as in the dataset. In Table~\ref{tab:metrics}, DeepTreeGAN is compared to other state-of-the-art GANs and found to be competitive in this context. Diffusion and flow-matching based models~\cite{buhmann2023epicly,Mikuni2023} have recently provided leading results for the JetNet dataset, but sample production is generally slower compared to GAN approaches. The DeepTreeGAN code and weights are available on GitHub.\footnote{\texttt{\url{https://github.com/DeGeSim/nips23DeepTreeGANv2}}}

%% file: parts/metricstable.tex
\begin{table*}[bt]
    \centering
    \caption{Comparison of the proposed DeepTreeGAN to EPiC-GAN~\cite{Buhmann2023} and MDMA~\cite{kaech2023mdma}. 
    The metrics were computed on a 25k event hold-out sample of the \JN 150 top quark dataset. The `Limit' row gives the measured in-sample distance by bootstrapping the hold-out dataset.  Lower is better for all metrics. Values taken from \cite[Table 1]{kaech2023mdma}. The metrics are provided by the \JN library~\cite{JetNetLib}. The best performance is printed in bold.} \label{tab:metrics}
    \begin{tabular}{llllll}
        \toprule
        Model & $W_1^M (\times 10^{3})$ & $W_1^P
(\times 10^{3})$ & $W_1^{EFP}(\times 10^{5})$ &$FPD(\times 10^{4})$ \\
        \midrule
Limit & $0.42 \pm 0.09$ & $0.12 \pm 0.04$ & $1.22 \pm 0.32$ &  $1.2 \pm 0.6$ \\
EPiC-GAN & $0.69 \pm 0.08$ & $0.65 \pm 0.03$ & $2.67 \pm 0.39$ &  $22 \pm 1$ \\
MDMA & $\mathbf{0.57 \pm 0.09}$ & $ 0.10 \pm 0.02$ & $\mathbf{2.12 \pm 0.64}$ & $5.3 \pm 0.9$ \\
DeepTreeGAN\hspace{-.5em} & $1.49\pm0.04$ & $0.13\pm0.02$ & $5.01\pm0.08$ & $\mathbf{3.4\pm0.7}$\\
        \bottomrule
    \end{tabular}
\end{table*}
% 'w1m': (1.490, 0.04585)
% 'w1p': (0.13, 0.015)
% 'w1efp': (5.01, 0.080)
% 'kpd': (-0.05760, 0.1478)
% 'fpd': (3.38, 0.65)
% 'fpnd': (0.1005, nan)
% {'w1m': (1.4903996480405333, 0.045855542477895916), 'w1p': (0.137529198161473, 0.0151996757641081), 'w1efp':
% (5.013644195369554, 0.0806091962241631), 'fpnd': (0.10054863947144099, nan), 'kpd': (-0.0057607056005437585,0.014783962010685103), 'fpd': (0.3384422204450645, 0.0650669371920478)}